\documentclass[aps,amsfonts,prx,twocolumn,showpacs]{revtex4-1}

\usepackage{subfigure}
\usepackage{wrapfig}
\usepackage{textcomp}
\usepackage{graphicx}
\usepackage{amssymb}
\usepackage{amsmath}
\usepackage{bm}
\usepackage{amsmath, amsthm, amssymb}
\usepackage{hyperref}

\usepackage{amsfonts}
\usepackage{dcolumn}
\usepackage{float}
\usepackage{bm}

\def\beq{\begin{equation}}
\def\eeq{\end{equation}}
\def\bea{\begin{eqnarray}}
\def\eea{\end{eqnarray}}

\newcommand{\lt}{<}
\newcommand{\gt}{>}

\begin{document}

\title{Weak crystallization theory of metallic alloys}

\author{Ivar Martin$^1$, Sarang Gopalakrishnan$^2$, and Eugene A. Demler$^2$}

\affiliation{$^1$ Materials Science Division, Argonne National Laboratory, Argonne, Illinois 60439, USA\\
$^2$ Department of Physics, Harvard University, Cambridge MA 02138, USA.
}

 \date{\today}

\begin{abstract}
We extend the Weak Crystallization theory to the case of metallic alloys. The additional ingredient -- itinerant electrons -- generates nontrivial dependence of free energy on the angles between ordering wave vectors of ionic density. That leads to stabilization of FCC, Rhombohedral, and icosahedral quasicrystalline (iQC) phases, which are absent in the generic theory with only local interactions. The condition for stability of iQC that we find, is consistent with  the Hume-Rothery rules known empirically for majority of stable iQC; namely, the length of the primary Bragg peak wavevector is approximately equal to the diameter of the Fermi surface.

\end{abstract}

\maketitle

{\em Introduction.}
Crystallization is probably the most familiar but one of the hardest to analyze  phase transitions. The workhorse of the theory of phase transitions, the Ginzburg-Landau theory \cite{CL}, cannot be easily applied to many of the crystallization transitions since they tend to be strongly first order, i.e., the order parameter experiences a large jump at the transition. 

Crystals are best characterized in  reciprocal space, where the onset of long-range order is signaled by the appearance of resolution-limited Bragg peaks. The intensity of the Bragg peaks reflects the density distribution in a material -- for smooth density modulations, as in the case of liquid crystals, only few harmonics of principal peaks located  on a momentum shell of radius $q_0$ are needed to fully describe the state. Then, the intensity of principal harmonics is the order parameter, and when it is small near the transition (relative to the average density), the application of  the Ginzburg-Landau theory is justified.
In atomic crystals, however, density is highly peaked at the equilibrium positions of atoms, and the number of relevant Bragg peak harmonics  scales in proportion to the ratio of the unit cell size  to the atomic size (smeared by thermal and quantum fluctuations).  In a typical crystal, the thermal fluctuations of atoms are $15-30 \%$ of the lattice spacing at the melting transition \cite{lindemann}; therefore, to  accuratley describe the transition, multiple harmonics of $q_0$ are required.  The appearance of strong modulation immedaitely at the phase transition, with multiple  Bragg peaks forming reciprocal lattice, is the signature of strongly first order transition.
A  special case of a crystalline solid is a {\em quasicrystal}, where atoms lack simple spatial periodicity; yet, in the reciprocal space, resolution-limited Bragg peaks appear in a self-similar arrangement  inconsistent with crystallographically allowed point-group symmetries \cite{Shechtman, Levine, DiV, Janot, Trebin}.

Weak Crystallization theory \cite{katz}  applies Ginzburg-Landau machinery to crystallization by assuming that only the Bragg peaks on a single momentum shell are significant enough to affect energy. While most naturally applicable to liquid crystals, the theory has been used to predict ubiquity of Body Centered Cubic (BCC) crystals near crystallization temperature \cite{AMT}, to study effects of fluctuations \cite{Braz}, and, with some modifications, to address the problem of stability of quasicrystals \cite{Bak, Levitov, Jaric, Mermin, Ho}. As such, it has been a useful symmetry-based tool to study the crystallization transition, even beyond its immediate range of validity. However, in its standard spatially-local form, weak crystallization theory is incapable of obtaining many of the experimentally relevant crystalline states, such as simple cubic, rhombohedral, or Face Centered Cubic (FCC), and the heuristic modifications of it have not been microscopically justified.

Crystals are often (qualitatively) separated into classes based on the dominant type of interaction that holds them together: ionic, covalent, molecular, and metallic. The crystal structure depends on a variety of details, such as ionic charge and electronic orbital structure, etc.  It is therefore remarkable, that in the case of metallic crystalline alloys simple empirical rules exist. These rules were identified by Hume-Rothery \cite{HR}  who has found that metallic alloys are particularly stable when in addition to the requirement that atoms be of similar size and electronegativity, the value of the average valence per atom (``$e/a$" ratio) be close to certain ``magic" values, which depend on the crystal structure. The optimal $e/a$ ratio has been argued to be associated with a particular geometrical matching condition, when the itinerant (nearly-free) electron Fermi surface ``just crosses" the boundary of the first Brillouin zone \cite{Jones}.
Regardless of interpretation, this observation points to an important, if not decisive, role that itinerant electrons play in determining the crystal structure. This is indeed not surprising given that electrons are an effective mediator of long-ranged and multi-body interionic interactions. 

Significantly, majority of stable quasicrystals have turned out to be Hume-Rothery alloys \cite{qHR}, i.e.,  they are stable for narrow ranges of $e/a$. Despite nominally large conduction electron concentration, their electrical and thermal conductivities are exceptionally low \cite{CpRho}, consistent with strong scattering around the Fermi surface. In analogy to regular crystals, attempts have been made to construct a theory of quasicrystals accounting for the  Hume-Rothery rules by perturbatively including electron scattering on quasiperiodic ionic potential \cite{Friedel}.  
Just as in the case of regular crystals, such approach is problematic (see Appendix \ref{sec:others}). 
There have also been attempts to understand the formation of quasicrystals in terms of Weak crystallization theory \cite{Levitov,Mermin,Jaric,Bak, Ho}; however, there have been no microscopic justification of the theoretical assumptions, and in particular these theories cannot account for the fact that most of quasicrystals follow the Hume-Rothery rules.

Here we extend the Weak Crystallization theory to metallic systems. We do so by explicitly introducing itinerant electrons that couple to the ionic density, and integrating them out. We find that interionic interactions generated by electrons qualitatively modify the generic weak crystallization theory, stabilizing FCC, rhombohedral, and, notably, icosahedral quasicrystal (iQC) states. The Hume-Rothery rules emerge from the interplay of two length scales -- the preferred interionic distance, $1/q_0$,  and the Fermi wavelength of itinerant electrons, $1/k_F$. In particular, for iQC we find $q_0 \approx 2k_F$, consistent with empirical observations \cite{qHR}. 

In our approach we explicitly calculate the electronic contributions to the quadratic, cubic, and quartic in density terms in the Ginzburg-Landau (GL)  energy. In this regard it is analogous the direct derivation of Free energy in the cases of superconductivity \cite{Gorkov} and charge density waves \cite{Norm}. 
Of crucial importance is the strong  (singular in the limit of zero temperature)  dependence of quartic term on the angle between the Bragg wave vectors. At some angles, which can be tuned by changing the size of the electronic Fermi surface, the repulsion between different Fourier components of ionic density can be reduced or even turned into attraction.
For a generic Fermi momentum $k_F$, this effect favors formation of rhombohedral states, with three ordering wavevectors with  identical angles $\alpha_{min}(k_F) $ between them.
For more finely tuned $k_F$, the minimum interaction can be reached at angles $\alpha_{min}$ (and equivalent $\pi-\alpha_{min}$ angles) that favor  simple cubic, FCC, or iQC.
It is interesting to note that FCC state additionally benefits from a {\em noncoplanar} fourth order interaction terms that can be always made attractive by appropriate choice of the order parameter signs, expanding the domain of its stability. 

{\em Weak crystallization theory and its extension to metals.} In what follows, we shall keep only momenta of length $q_0$; i.e., we shall make the ansatz $\rho(\mathbf{x}) = \rho_0 + \sum\nolimits_{\mathbf{|k| = q_0}} \mathrm{Re} [ \rho_{\mathbf{k}} e^{i \mathbf{k \cdot x}}]$ for ionic density. As discussed above, this ansatz, which is central to ``weak crystallization'' theory~\cite{katz}, is strictly valid only where the crystallization transition is weakly first-order and its latent heat is small. Outside this regime, our results will not be quantitatively accurate; nevertheless, we expect them to provide guidance as to what kinds of crystal structures are favored.

We proceed by writing down a general Ginzburg-Landau (GL) free energy functional, ${\cal F} = F_0 + H_0 + V$, where

\begin{eqnarray}
F_0 & = & \sum\nolimits_\mathbf{q} r(q) |\rho_\mathbf{q}|^2 + \frac{\lambda_3} {3!}\sum\nolimits_{\mathbf{q}_i} \rho_{q_1}\rho_{q_2}\rho_{q_3}\delta(\sum{q_i}) \nonumber \\
& & \quad + \frac {\lambda_4}{4!}  \sum\nolimits_{\mathbf{q}_i} \rho_{q_1}\rho_{q_2}\rho_{q_3}\rho_{q_4}\delta(\sum{q_i}) \label{eq:WC}\\
H_0 & = & \sum\nolimits_\mathbf{k} [E(k) - \mu] c^\dagger_{\mathbf{k}} c_{\mathbf{k}} \\
V & = & \sum\nolimits_{\mathbf{kq}} v(q) \rho_{q} c^\dagger_{k} c_{k - q}
\end{eqnarray}

\begin{figure}
\includegraphics[width=1\columnwidth]{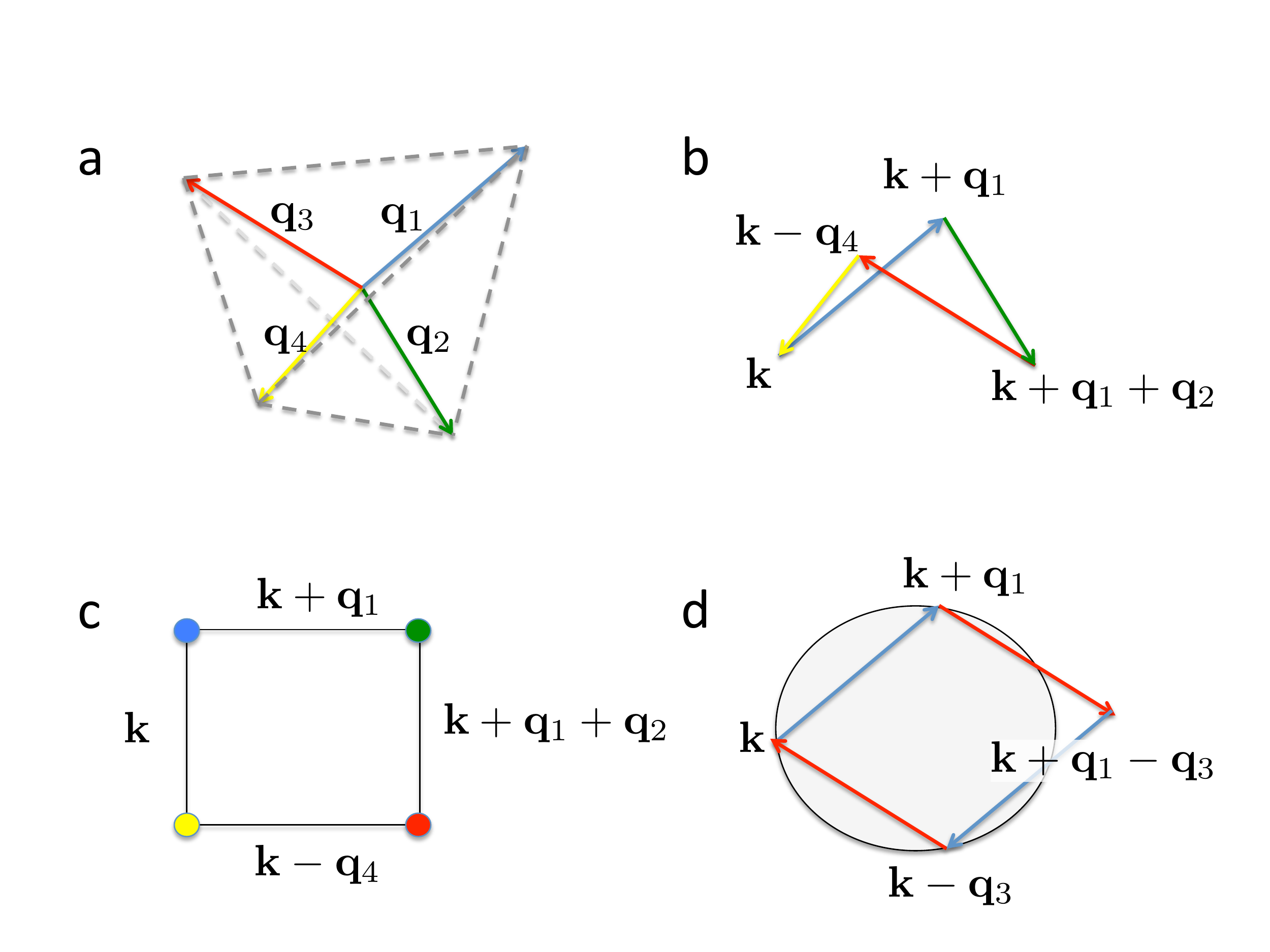}
\caption{(a) First shell of reciprocal lattice vectors for the case of FCC real space (BCC reciprocal) lattice vectors. Four vectors point from the center of a tetrahedron to its vertices. (b) Example of a scattering path between electronic momentum states induced by the ionic densities modulated at wavevectors in (a). This process makes contribution to $w(\{{\mathbf q}_i\})$ in Eq. (\ref{eq:F}). (c) Feynman ``box" diagram for the fourth order contribution to the GL free energy. Lines correspond to electronic Green functions; vertices to ionic densities with a given wavevector. (d) An example of coplanar contribution to free energy that only contains two pairs of $\pm \mathbf q_i$ [a contribution to $u(q_0, \alpha_{ij})$]. As shown, corresponds to the ``resonance" conditions satisfied: tree or more electronic momenta are on a great circe of the Fermi surface.}\label{fig:diag}
\end{figure}

Here, $F_0$ describes the physics of ions and {\em core} electrons in the absence of itinerant electrons. The minimal (``local") assumption that is commonly made is that interactions $\lambda_3$ and $\lambda_4$ are simply constants. However, as interatomic interactions set a preferred length-scale for crystallization even in the absence of conduction electrons, this length-scale is introduced into the second-order term, via the weak-crystallization form $r(q) = r_0 + \chi (|q| - q_0)^2$. As already stated, we will restrict our attention to density modes that are precisely at $q = q_0$ \cite{cHo}. The second term, $H_{0}$, describes the itinerant electrons: for simplicity we shall treat these as noninteracting. The third term, $V$, describes the interaction between itinerant electrons and atoms. As we are only concerned with density modulations satisfying $q = q_0$, and the interaction is assumed to be spherically symmetric, we can parameterize the interaction strength entirely by its Fourier component at momentum transfer $q_0$, viz. $v \equiv v(q_0)$. Thus we need not make any assumptions about screening of the Coulomb interaction. The kinematic constraints $\sum\nolimits_{\mathbf{q}_i} $, in combination with the $q = q_0$ restriction, strongly limit the number of allowed terms. Namely, the cubic term is only non-zero for triplets of $\mathbf{q}_i$ forming equilateral triangles, and thus favors hexagonal and BCC crystal structures \cite{AMT}.
The quartic term obtains generically from combining $\pm \mathbf{q}_i$ with $\pm \mathbf{q}_j$. It can also appear in the situation when four $\mathbf{q}_i$ form a non-copanar quadrilateral [e.g., the
geometry in Fig. \ref{fig:diag}(a)].

\begin{figure}
\includegraphics[width=0.8\columnwidth, angle = 90]{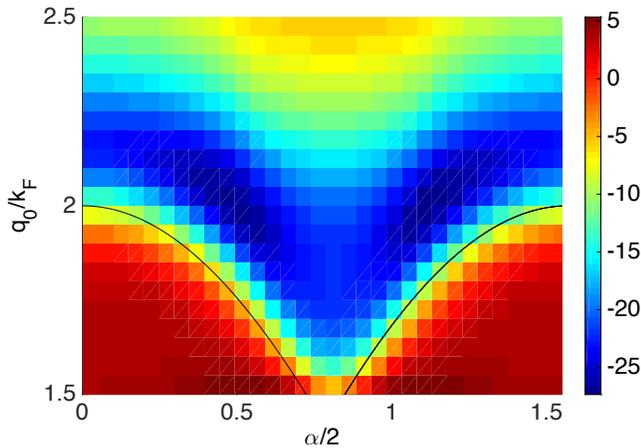}
\caption{Electronic contribution $u(\alpha)$ to 4th order term in Ginzburg-Landau energy functional Eq. ~(\ref{eq:F}). Temperature is $T = 0.1 E_F$. Black lines mark location of zero-temperature singularity, $Q/k_F = 2\cos\frac{\alpha}{2}$. }\label{fig:vert1}
\end{figure}

{\em Electronic contribution to Weak crystallization energy functional.}
We now integrate out the conduction electrons to arrive at a description that is purely in terms of the atomic densities. The procedure is analogous to the derivation of Ginzburg-Landau functional for superconductivity or a charge density wave states.  The difference is that the ionic density order parameter, a priori, can have an arbitrary number of components, and the energy functional should be able to predict not only the magnitude of the order parameter, but also the number and orientation of its components.  The latter determines the type of crystalline state.

{As the free energy functional $\mathcal{F} = F_0 + H_0 + V$ is quadratic in fermion operators, we can integrate out the fermions; this allows us to write the partition function purely in terms of ionic densities, as $\mathcal{Z} = \exp[- \beta (F_0 - \Delta F)]$ where $F_0$ is defined in Eq.~\eqref{eq:WC} and $\Delta F$ is given by the following perturbation series, which we have resummed using the linked cluster theorem}:

$$\Delta F = -\frac{1}{\beta}\sum_n \frac{(-1)^n}{n}\int d\tau_1 ... d\tau_n \langle {\cal T}_\tau V(\tau_1) ... V(\tau_n)\rangle_{conn},$$
Explicit expressions are given in Appendix \ref{sec:det}. We now expand $\Delta F$ to quartic order in the bosonic densities; this yields, for the free energy functional $F \equiv F_0 + \Delta F$:

\begin{widetext}
\begin{eqnarray}
F  &=&  \sum\nolimits_\mathbf{q_i} \tilde r(q) |\rho_\mathbf{q}|^2 + \tilde \lambda_3(q_0) \sum\nolimits_{\triangle} \rho_{\mathbf q_1}\rho_{\mathbf q_2}\rho_{\mathbf q_3}\delta(\sum{\mathbf q_i}) \nonumber\\
&&+ \frac 12 \sum\nolimits_{\mathbf{q}_i\ne \mathbf{q}_j }[\lambda_4 + u(\alpha_{ij})]  |\rho_{\mathbf q_i}|^2 |\rho_{\mathbf q_j}|^2  
+  \frac 14\sum\nolimits_{\mathbf{q}_i}[\lambda_4 + u(0)]  |\rho_{\mathbf q_i}|^4  + 
\sum\nolimits_\Box [\lambda_4 + w(\{\mathbf q_i\})]\rho_{\mathbf q_1}\rho_{\mathbf q_2}\rho_{\mathbf q_3}\rho_{\mathbf q_4}\label{eq:F}
\end{eqnarray}
\end{widetext}

The symbols $\sum\nolimits_{\triangle}$ and $\sum\nolimits_\Box$ indicate summation over unique triangles and non-planar quadrilaterals of $\mathbf{q_i}$; $\alpha_{ij}$ is the angle between vectors $\mathbf q_i$ and $\mathbf q_j$.

{\em Numerical results. } Figure \ref{fig:vert1} shows  $u ( \alpha)$  for various values of $q_0$ at $T = 0.1 E_F$. Already at this not very low temperature certain features become apparent. For $q_0/k_F \sim \sqrt{2}$, a minimum in $u(\alpha)$ develops around $\alpha = \pi/2$, which then splits into two minima for larger values of $q_0$. In the limit of zero temperature, a singularity develops along the line $q_0/k_F = 2\cos\frac\alpha 2$. Geometrically this condition corresponds to the configuration when three momenta connected by scattering off the ionic order parameter, $\mathbf  k$, $\mathbf  k + \mathbf q_1$ and $\mathbf  k + \mathbf q_2$ can all simultaneously be on a great circle of the Fermi surface (Fig. \ref{fig:diag}d). Near this line, the vertex is repulsively divergent for smaller $q_0$ and attractively divergent for larger $q_0$ as $T \to 0$. This singular behavior is a four particle/hole analog of the particle-hole divergence in 1D that drives Peierls instability. Naturally, such a strong angular dependence of $u(\alpha)$ at temperatures much lower than $E_F$ can influence the energetic balance between different crystalline phases. It should be noted, that the angular dependence is a result of sharply defined Fermi surface; and temperatures comparable or higher than the Fermi energy it becomes smeared out.

\begin{figure}
\includegraphics[width=1\columnwidth]{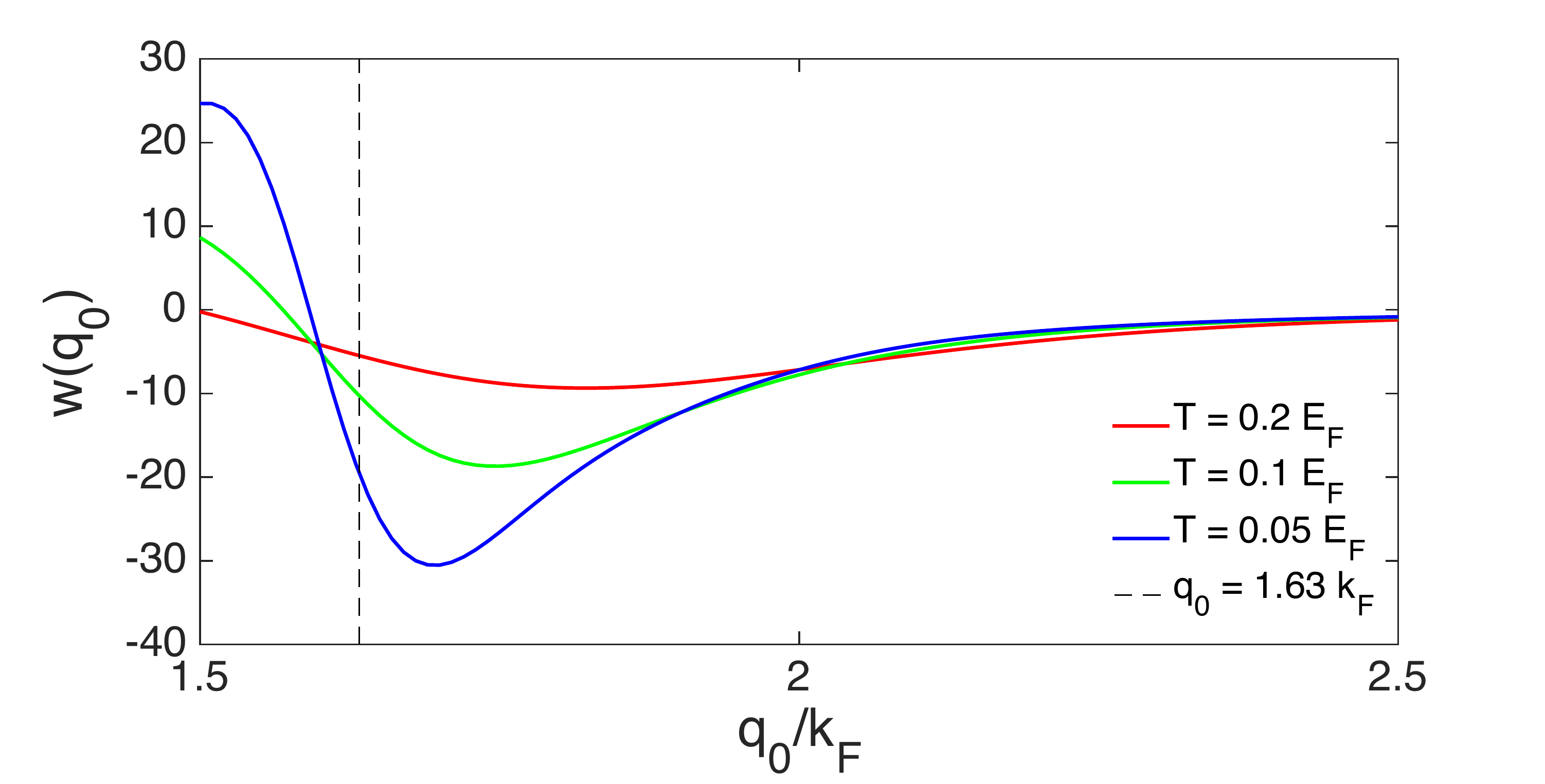}
\caption{Temperature and $q_0$ dependence of noncoplanar electronic contribution $w(q_0)$ to 4th order term in Ginzburg-Landau energy for FCC crystal. Black vertical line marks location of zero-temperature singularity, $q_0/k_F = 2\cos\frac{\alpha}{2}$, where $\alpha = \arccos(1/3)$. }\label{fig:vert2}
\end{figure}

The non-coplanar terms $w(\{\mathbf q_i\})$ are less generic than $u(\alpha)$ since they require four distinct wavevectors to add up to zero. A case where these terms are important is the FCC crystal, whose first Bragg shell (the set of shortest symmetry-related reciprocal lattice vectors) is comprised of eight vertices of a cube; hence there are two non-trivial quadruplets of wavevectors  that correspond to the vertices of two tetrahedra (see Fig. \ref{fig:diag}). The significance of non-coplanar terms is that they can always be made to lower energy by appropriate choice of the relative signs of constituent $\rho_{\mathbf q_i}$. For the momentum-independent interactions this does not change the fact that stripe state has the lowest energy within Weak Crystallization theory. However, inclusion of the electron-induced interaction can change he situation dramatically. In Figure \ref{fig:vert2} we plot $w(\{\mathbf q_i\})$ as a function of $q_0$ for FCC. We find that it has features similar to $u(\alpha)$; namely, when $q_0/k_F = 2\cos\frac{\alpha_t} 2$, with $\alpha_t = \arccos(1/3)$ the tetrahedral angle, $w$ diverges as $T\to 0$. This enhanced interaction is the cause of a large region of stability of FCC phase that we find.

{\em Phase diagram.}
To construct the phase diagram, we  first consider the set of variational states that contain \(N\) pairs of \(\pm \mathbf q_i\), where
all \(\mathbf q_i\)'s are symmetry related, and hence have exactly the same set
of neighbors. Then, all the Fourier amplitudes are identical, \(|\rho_{\mathbf q_i}| = \rho\) and the free energy is

\[F = N r_0|\rho|^2 +\frac N 2 \left[\sum_{j\ne 0} \tilde u (\alpha_{0j}) + \frac{\tilde u_0}{2}- \frac {2M_{\square}} N |\tilde w(\{{\bf q}_i\})|\right ]|\rho|^4 ,\]
where we redefined $\tilde u = u + \lambda_4$ and $\tilde w = w + \lambda_4$ for compactness. We assumed that all $M_{\square}$ quadruplets have the same $w(\{{\bf q}_i\})$ (the case for FCC) and that vectors \(q_i\) do not form equilateral triangles, and hence the cubic invariant that could stabilize BCC (FCC reciprocal) crystal does not contribute (the latter assumption should become valid for sufficiently large negative $r_0$). Now it only remains to minimize the energy to obtain,

\[ |\rho|^2 = -\frac{r_0}{\sum_{j\ne 0} \tilde u (\alpha_{0j}) + \frac{\tilde u_0}{2} - \frac {2M_{\square}} N |\tilde w(\{{\bf q}_i\})|}\]

and
\beq
F =   -\frac{r_0^2}{\frac 2 N\sum_{j\ne 0} \tilde u (\alpha_{0j}) + \frac{\tilde u_0}{N}- \frac {4M_{\square}} {N^2} |\tilde w(\{{\bf q}_i\})|}.\label{eq:Var}
\eeq

It is important to note that the pure electronic vertex $u$ is negative (attractive) in a wide range of $q_0$ and $\alpha$, which taken by itself could cause an absolute instability. In this regime, one cannot truncate $F$ at fourth order, but must include higher-order terms in the GL expansion to find stable equilibrium states. However, the structureless local interaction $\lambda_4$  restores stability while maintaining the strong angle-dependence of the interactions.

\begin{figure}
\includegraphics[width=1\columnwidth]{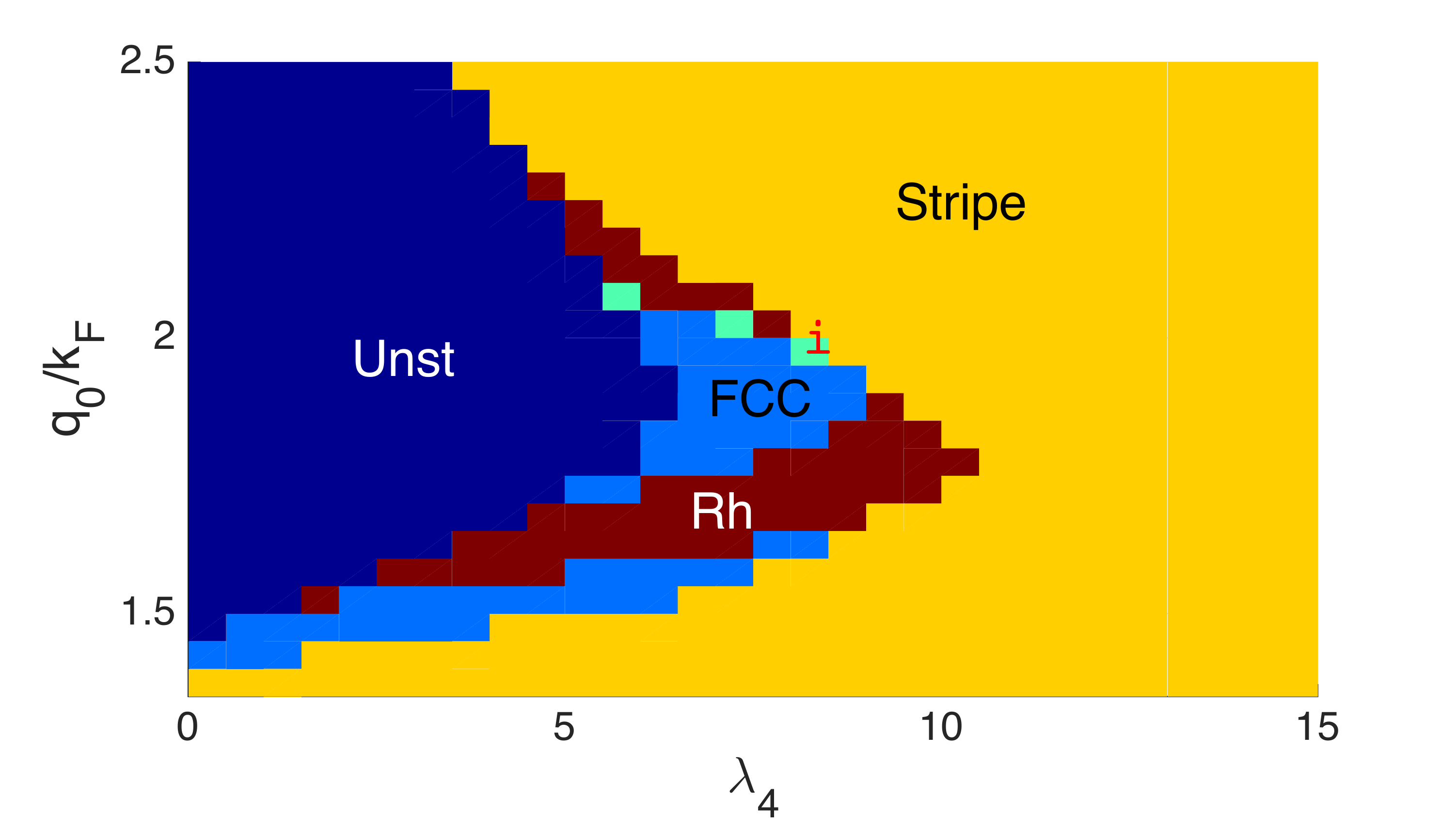}
\includegraphics[width=1\columnwidth]{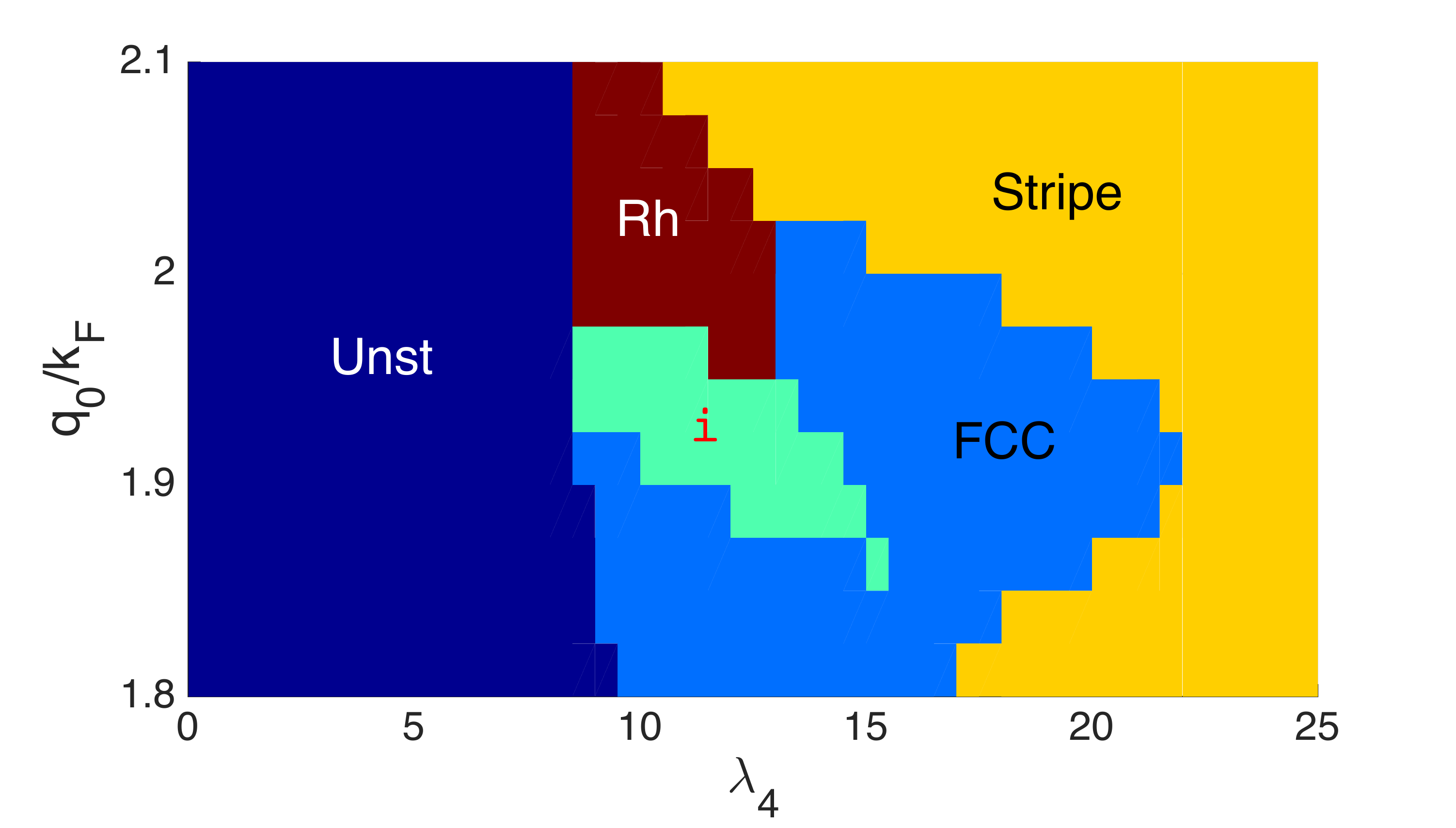}
\includegraphics[width=1\columnwidth]{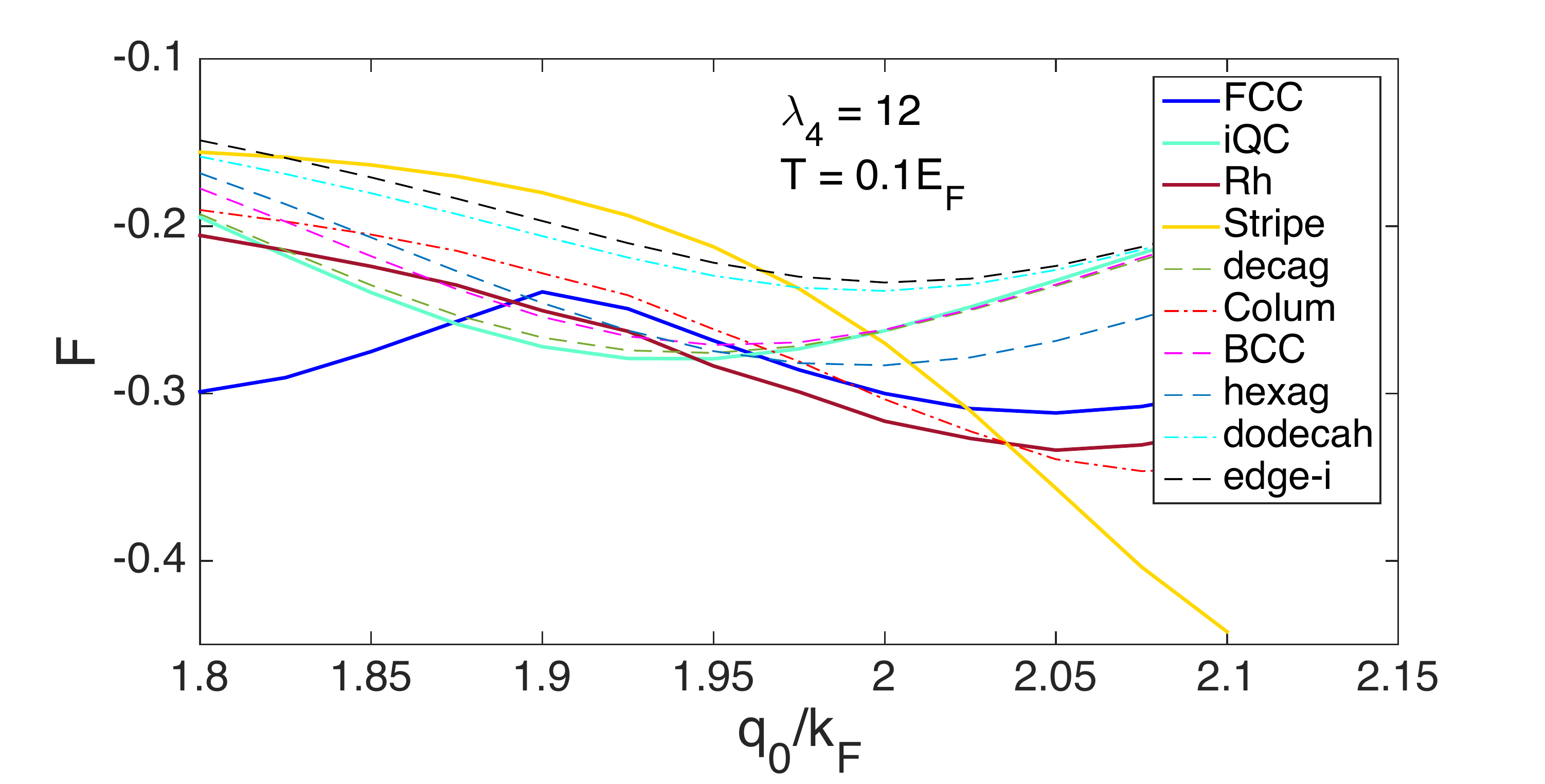}
\caption{Variational phase diagram for $T = 0.2 E_F$ (top) and $T = 0.1 E_F$ (middle) as a function of local repulsion strength $\lambda_4$ and the  ionic ordering vector $q_0$. At small values of $\lambda_4$ (region marked Unst), the fourth order terms in GL for one or more of the variational state becomes negative, signaling the need to consider higher order stabilizing terms. Lower panel shows Energy comparison of different variational states for a fixed value of $\lambda_4$ at $T = 0.1E_F$.}\label{fig:pd}
\end{figure}

The results of our analysis are presented in Figure~\ref{fig:pd}. We find only four stable phases: rhombohedral, striped, FCC, and iQC (i.e., icosahedral quasicrystal). The other symmetric variational states we explored are always higher in free energy than these (see lower panel of Figure \ref{fig:pd} for energy comparison and Appendix \ref{sec:var} for details of the variational states). The overall shape of the phase diagram can be understood as follows. When the structureless interaction $\lambda_4$ is absent or too weak, as noted above, the free energy can become unstable at quartic order. On the other hand, when $\lambda_4$ is dominant, the electron-induced interaction can be ignored, and we recover the standard weak-crystallization result that the equilibrium state is striped (or smectic). When we are far from the matching condition $q_0 \sim 2 k_F$, or the temperature is relatively high, the interactions are not strongly angle-dependent, and these are the two dominant possibilites. On the other hand, when the structureless and electronic contributions are of similar magnitude \emph{and} the temperature is low, the angle-dependence of the electron-mediated interaction stabilizes nontrivial crystalline phases.

The most significant qualitative feature of the phase diagram for intermediate values of $\lambda_4$ is the dominance of FCC and rhombohedral phases, with balance shifting in favor of FCC at lower temperatures. The reason for this trend is that FCC has two appealing features -- (1) it has only one inter-$\mathbf q$ angle $\alpha_{ij}$  (up to $\pi - \alpha_{ij}$), and (2) it has lower energy due to the presence of non-coplanar 4th order terms. Rhombohedral crystal only has former feature, and thus only becomes competitive when the angle $\alpha_{min}$ that minimizes $u(\alpha)$ is sufficiently far from the tetrahedral angle. Surrounded by the FCC and rhombohedral phases is the iQC phase.  The key advantage of iQC phase is that it has large number (six) of $\pm \mathbf q_i$ pairs, all separated  by the same angle $\alpha_i =  2 \sin^{-1} (\gamma^2 +1)^{-1}\approx 63.43^o$ ($\gamma$ is the Golden mean). Even thought iQC can not benefit from the non-coplanar energy terms, when the optimal angle $\alpha_{min}$ is close to the $\alpha_i$, iQC can beat both FCC and Rhombohedral. Finally, for large $\lambda_4$ we recover the Stripe phase predicted by the original featureless Weak crystallization theory.

Apart from the FCC phase with its non-coplanar terms in energy, the  phase diagram can be understood as follows. For the states with only one non-trivial inter-$\mathbf q$ angle $\alpha_{min}$,  the denominator in Eq. (\ref{eq:Var}) is
$\tilde u(\alpha_{min}) + [\tilde u(0) - 2\tilde u(\alpha_{min})]/N$. Thus, if the second term is positive, it favors large $N$; it it is negative, 
then $N = 1$. Note that for $u(0) = 2 u(\tilde\alpha)$ states with all possible $N$'s are energetically degenerate (can be seen in triple crossing point in Figure \ref{fig:pd}(lower) at $q_0/k_F \approx 2.05$).

In construction of the phase diagram we have only considered highly symmetric states. We now discuss possible deviations from these assumptions. First, we can ask whether the highly symmetric crystal states are stable with respect to ``Bragg-fractionalization," namely whether it may be beneficial to split Bragg peaks into multiple nearby ones.  From the fact that $\tilde u(\alpha\to 0 ) = \tilde u_0$, which can be explicitly demonstrated for electron-mediated and local interaction, for $\tilde u_0>0$, lack of fractionalization follows trivially (see Appendix \ref{sec:frac}). The next possibility is a distortion of peaks from symmetric positions. Clearly this is not a concern for Rhombohedral state, but could be for iQC and FCC. 
Here we specifically ask whether iQC will remain stable even if $ \alpha_{min}$ is not exactly $\alpha_i$ (see Appendix \ref{sec:dist}). Due to its high symmetry iQC cannot naturally distort, unlike, e.g., Rhombohedral state.
To answer this question, we have expanded the interaction energy around the symmetric iQC state. We have found that if $u'(\alpha _i) \lt - (2/3) u''(\alpha _i)$, then iQC spontaneously distorts into a lower symmetry state, i.e. a distortion could occur if $\alpha_{min} > \alpha_i$ (``compresses springs"). This criterion also shows that if $u(\alpha)$ is sufficiently smooth, as it is for temperatures not very much smaller than the Fermi energy, then undistorted iQC should in fact be quite stable. Indeed, expanding around $\tilde\alpha$, we find that criterion for instability is $ \tilde\alpha_{min} - \alpha_i  > 2/3$, i.e., the minimum is at least 40$^o$ away (above) from  the icosahedral angle.
This estimate is based on the assumption of smoothness of $u(\alpha)$, which is violated at temperatures much below the Fermi temperature. Thus, for quasicrystals that form under such conditions, there is a possibility of distorted iQC, as well as a structural transition from perfect to distorted iQC as a function of temperature.

We have also explored possible ordered states in an alternative fashion by applying unconstrained stochastic minimization of the free energy functional. To simplify simulations, we neglected the cubic and non-coplanar quartic terms and thus cannot fully capture FCC and BCC phases; however,  the advantage of this method is that it provides an unbiased treatment for arbitrary multi-$q$ states that are not required to possess any special symmetries. We start from random configuration of several hundred components $\rho_{\mathbf q_i}$ with $\mathbf q_i$ on a sphere or radius $q_0$.
We then iteratively minimize  energy by randomly selecting $\rho_{\mathbf q_i}$ and changing its value and position on $q_0$ sphere in the direction of decreasing energy. The minimization results are consistent with the variational phase diagram in Figure \ref{fig:pd} (modulo underestimating the stability of FCC).  Due to the stochastic nature of the algorithm, however, it sometimes converges to other states. In particular, in the region of stability of iQC, the final state is rather commonly the decagonal state, which is approximately the iQC state with one pair $\pm q_i$ removed.  This state is and example of a 2d quasicrystal -- it is periodic along one axis and quasiperiodic in the plane perpendicular to it. Even though the energy of this state is very close to the iQC, we have not observed it ever to be lower in energy than the perfect iQC (consistently with Figure \ref{fig:pd} (lower)). The energy difference is nevertheless sufficiently delicate, so one cannot rule out that for modified conditions decagonal state may appear as the lowest energy state in the phase diagram.

The conjecture that stability of 3D quasicrystals is associated with  ``bond-orientational order" that favors specific inter-$\mathbf q_{i}$ angles within Weak Crystallization theory has been previously expressed by   Mermin and Troian \cite{Mermin} and Jaric \cite{Jaric}. In Ref. \cite{Mermin} 
an auxiliary field was introduced to generate preferred  inter-$\mathbf q_{i}$ angle, however, no physical justification was given as to the nature of this field. The key result of our work is that itinerant electrons play the role similar to the auxiliary field postulated in \cite{Mermin}. On the experimental side, it has been found that the optimal e/a ratio observed in quasicrystals corresponds to the approximate matching between the quasisrystalline {\em quasi}-Brillouin zone and the electronic Fermi surface, that is, the length of the dominant Bragg wave vector equals the diameter of the Fermi surface, $2k_F$.  This is indeed what we find.

In conclusion, we have analyzed the effects of electron-ion interactions on crystallization transition within Weak Crystallization theory. We found that the angular-depended multi-ion interactions induced by electrons can lead to stabilization of such empirically common but elusive, within the standard theory, states as Rhombohedral, FCC, and icosahedral quaiscrystals. The stability conditions can be recast in terms of Hume-Rothery rules connecting primary ionic ordering wave-vectors and the size of the electronic Fermi surface. Our results are obtained within the assumption that the cubic invariants are less relevant than the quartic ones, i.e., at temperatures sufficiently lower than the temperature of mean field transition ($r_0 = 0$). Near the transition, more careful analysis of fluctuations is required \cite{Braz}, which will be the subject of future work.

{\em Acknowledgements}  Authors would like to thank P. Goldbart, Z. Nussinov, D. Levine, P. Steinhardt, and A. Rosch for useful discussions. IM acknowledges support from Department of Energy, Office of Basic Energy Science, Materials Science and Engineering Division. SG and EAD acknowledge support from Harvard-MIT CUA, NSF Grant No. DMR-07-05472, AFOSR Quantum Simulation MURI, the ARO-MURI on Atomtronics, ARO MURI Quism program.

%

\appendix

\section{Other approaches to crystallization theory and their limitations.}\label{sec:others}

The most common approach to determine the lowest-energy crystal structure is based on variants of microscopic density functional theory, which specifies atoms with their electronic shells and attempts to optimize their spatial arrangement \cite{dft}. Due to computational complexity, this is a variational approach that can effectively treat only periodic arrangements of atoms. Near the melting transition, application of this method becomes difficult since atoms in a liquid lack spatial periodicity. In that regime, methods combining density functional theory with molecular dynamics are applied, but only with limited success \cite{dft2}. 

Periodic {\em approximants} to quasicrystals have also been studied by density functional theory \cite{abin}; application of this method, however,   requires a very large number of atoms to be explicitly considered and optimized for the approximants energies to provide a good estimate for quasicrystals, even away from the melting transition.

Another, semi-microscopic, approach is based on the Peierls instability-type arguments. There, one studies the features in the electronic susceptibility and attempts to use its anomalies as a predictor of stable phases. This approach is problematic in the case of 3D alloys, as can be easily seen. We would like to have an unbiased predictor of an ordered state; therefore, the only starting point possible is free electron Fermi sea coupled to featureless (constant) ionic density. In 1D, electronic susceptibility diverges at $2k_F$  at $T = 0$, which leads to a density instability at this wavevector; this is the origin of charge density waves in many quasi-1D materials. In contrast, in 3D, free electron susceptibility is maximized at zero momentum and at $2k_F$ only has infinite first derivative (the cause of Friedel oscillations of electron mediated interaction \cite{RKKY}). However, this is insufficient to cause instability in ionic density -- theory would predict that the instability should occur at zero wavevector, i.e., at uniform density. Moreover, the  (quadratic) term in the GL theory that is proportional to the electronic susceptibility only includes a single density modulation, and thus cannot discern between orderings that contain multiple wave vectors. 

A way to go beyond the quadratic energy approximation is to include the ionic modulation non-perturbatively in electron dispersion \cite{Jones}. It has been argued this way that for a given crystal structure, the electronic energy is minimized when the Fermi surface ``just crosses" the Brillouin zone boundary. This naturally corresponds to crystal-specific optimal $e/a$ ratios, and thus appears to be consistent with the empirical Hume-Rothery rules.
Application of this approach to discriminate between energies of different crystalline and quasicrystalline states is, however, problematic, as it presupposes the knowledge of the amplitude of the periodic lattice (pseudo-)potential, which is different for different crystals. Since the energies of various states are typically rather similar, the uncertainty in the potential makes such approach unreliable.

\section{Details of electronic corrections to Free energy}\label{sec:det}

Integration of electronic degrees of freedom leads to the following corrections to the ionic free energy functional:

\begin{widetext}
\bea
\Delta F^{(2)} &=&  -\frac{1}{2 \beta} \int d\tau_1 d\tau_2 \langle {\cal T}_\tau \sum_{k_1,q_1, k_2,q_2}v_{q_1} \rho_{q_1} c^\dagger_{k_1+q_1}c{k_1}|_{\tau_1} v_{q_2} \rho_{q_2} c^\dagger_{k_2+q_2}c{k_2}|_{\tau_2}\rangle_{conn} \\
&=& \frac{|v_q \rho_{q}|^2}{2 \beta} \int d\tau_1 d\tau_2 G_p(\tau_2 - \tau_1) G_{p-q}(\tau_1 - \tau_2) =  \frac{|v_q \rho_{q}|^2}{2 \beta} \sum_{\omega_n, p} G_p(\omega_n) G_{p-q}(\omega_n),\\
\Delta F^{(3)} 
& =&  -\frac{v_{q_0}^3 \rho_{q_1} \rho_{q_2}\rho_{q_3} \delta(\sum{q_i})}{3 \beta} \sum_{\omega_n, p} G_p(\omega_n) G_{p - q_1}(\omega_n)G_{p - q_1 - q_2}(\omega_n),\\
\Delta F^{(4)} 
& =& \frac{v_{q_0}^4 \rho_{q_1} \rho_{q_2}\rho_{q_3}\rho_{q_4}\delta(\sum{q_i})}{4 \beta} \sum_{\omega_n, k} G_p(\omega_n) G_{p - q_1}(\omega_n)G_{p - q_1 - q_2}(\omega_n) G_{p - q_1 - q_2 - q_3}(\omega_n).
\eea
\end{widetext}

Here, $G_p(\omega_n) = (i\omega_n - \epsilon_p)^{-1}$.

As one can see, the new terms in the Ginzburg-Landau functional have the form similar to those already contained in $F_0$ (Eq. (\ref{eq:WC})). The second order term $\Delta F^{(2)}$ only serves to redefine $q_0$ and hence will be of no interest to us. The prefactor of the cubic term becomes a function of $q_0$. We find by numerical integration that the electronic contribution to this term is non-singular in the limit of zero temperature, and thus it does not introduce any qualitatively new features relative to those already in $F_0$.

The 4th order correction $\Delta F^{(4)} $ can diverge if certain geometric conditions are satisfied (see Figure 1 of the main text). Thus we concentrate here on this term only.

The first type of 4th order term that we will consider is self-interaction.  Self-interaction is generated by box diagrams with momentum transfers $(\mathbf q_1, \mathbf q_1, -\mathbf q_1, -\mathbf q_1)$  (type A) and $(\mathbf q_1, -\mathbf q_1, \mathbf q_1, -\mathbf q_1)$ (type B). 
The combinatorial multiplicities of these diagrams can be calculated as follows. At every vertex of the box diagram we can place $\rho_{\pm \mathbf q_1}$. For A type, there the sign pattern has to be such that same signs are adjacent, while for B -- interlaced. There are 4 ways to place two adjacent ++ in 4 boxes. $(++--),\ (-++-),\ (--++),\ (+--+)$. Hence A type has multiplicity 4. For B type, there are only two distinct ways to arrange: $(+-+-),\ (-+-+)$, and multiplicity is 2. Therefore, the self-interaction goes as 

$$\sum_i(4A + 2B)|\rho_{q_i}|^4.$$

There are two types of mutual interaction diagrams: those that contain only two pairs of $\pm \mathbf q_i$ (``coplanar" diagrams), and those that contain four distinct $\mathbf q_i$'s (``non-coplanar" diagrams). Coplanar diagrams depend only on one angle between $\mathbf q_1$ and $\mathbf q_2$ (for $\alpha = 0$ we get self-interaction).
There are three distinct contributions to $\Delta F^{(4)}$, which come from the following arrangements or momenta around the box diagram:  $\Delta F_1^{(4)}:(\mathbf q_1, -\mathbf q_1, \mathbf q_2, -\mathbf q_2)$ (type V$_1$),  $\Delta F_2^{(4)}:(\mathbf q_1, -\mathbf q_1, -\mathbf q_2, \mathbf q_2)$ (type V$_2$), and $\Delta F_3^{(4)}: (\mathbf q_1, \mathbf q_2, -\mathbf q_1, -\mathbf q_2)$ (type D).  Their combinatorial multiplicities are as follows:

V$_1$ + V$_2$: 4 ways to place $\mathbf q_1$, 2 ways to place $-\mathbf q_1$ next to it (PBC), 2 way to place  $\pm \mathbf q_2$: total 16. Hence there are 8 diagrams of each type.

D: 4 ways to place $\mathbf q_1$, 2 ways to place $\pm \mathbf q_2$. Total multiplicity is 8.

Therefore, the mutual interaction term is  

\bea
&\sum_{i\lt j}(8V_1 + 8V_2 + 8D)|\rho_{\mathbf q_i}|^2|\rho_{\mathbf q_j}|^2 \\
&= \sum_{i\ne j}(4V_1 + 4V_2 + 4D)|\rho_{\mathbf q_i}|^2|\rho_{\mathbf q_j}|^2.
\eea

Notice, that in the limit $\mathbf q_i \to \mathbf q_j$, $V_1 \to B$, and $(V_2, D)\to A$. Hence, going back to the original notation in terms of $u(\alpha)$ , we find that the full 4th order GL term is
$$\delta F^{(4)} = \frac12\sum_{i\ne j}u(\alpha_{ij})|\rho_{\mathbf q_i}|^2|\rho_{\mathbf q_j}|^2  + \frac 14 \sum_{i}u_0|\rho_{\mathbf q_i}|^2 $$
where $u(\alpha ) = 8(V_1 + V_2 + D)$ and $u_0 = u(\alpha = 0)$ (it can be shown explicitly that the limit $\alpha \to 0$ is continuous at finite temperature).

The interaction functions can be obtained by a mixture of analytical and numerical integration. The frequency summations could be performed with the help of contour integration,

\begin{widetext}

\bea
\{V_1,V_2\} 
&=& \frac{v_{q_0}^4 |\rho_{q_1}|^2 |\rho_{q_2}|^2}{4 } \sum_{ k}  \frac{n_F(\epsilon_1)}{(\epsilon_1 - \epsilon_2) ^2(\epsilon_1 - \epsilon_4)}+ \frac{n_F(\epsilon_4)}{(\epsilon_4 - \epsilon_1) (\epsilon_4 - \epsilon_2)^2} \nonumber\\
&& - \frac{n_F(\epsilon_2)}{(\epsilon_2 - \epsilon_1)^2 (\epsilon_2 - \epsilon_4)} - \frac{n_F(\epsilon_2)}{(\epsilon_2 - \epsilon_1)(\epsilon_2 - \epsilon_4)^2}      +  -\frac{n_F'(\epsilon_2)}{(\epsilon_2 - \epsilon_1)(\epsilon_2 - \epsilon_4)} ,
\eea
and 
\bea
D 
&=& \frac{v_{q_0}^4  |\rho_{q_1}|^2 |\rho_{q_2}|^2}{4 } \sum_{ k}  \frac{n_F(\epsilon_1)}{(\epsilon_1 - \epsilon_2) (\epsilon_1 - \epsilon_3)(\epsilon_1 - \epsilon_4)}+ \frac{n_F(\epsilon_2)}{(\epsilon_2 - \epsilon_1) (\epsilon_2 - \epsilon_3)(\epsilon_2 - \epsilon_4)} \nonumber\\
&& + \frac{n_F(\epsilon_3)}{(\epsilon_3 - \epsilon_1) (\epsilon_3 - \epsilon_2)(\epsilon_3 - \epsilon_4)} + \frac{n_F(\epsilon_4)}{(\epsilon_4 - \epsilon_1) (\epsilon_4 - \epsilon_2)(\epsilon_4 - \epsilon_3)}.
\eea

\end{widetext}

The numerical integration over momenta has to be done with care due to singular denominators. We found that numerical integration performs the best using the above forms of $\Delta F^{(4)}$ after introducing regularization  $\frac{1}{(\epsilon_3 - \epsilon_1)}\to Re \frac{1}{(\epsilon_3 - \epsilon_1 + i\Gamma)}$, with $\Gamma = 10^{-15}$   using Matlab 3d integration routine. For details and checks see \cite{knowen}.

In 3D, there is a possibility of a non-coplanar interaction diagrams. They obtain if there are non-trivial  $\{\mathbf q_1, ..., \mathbf q_4\}$ that add up to 0. Such diagrams exist for example for FCC lattice, which has reciprocal BCC.  There are 8 BCC reciprocal vectors, which can be split into two distinct quadruplets (tetrahedra). Each has total $4! = 24$ multiplicity. In the case of FCC, by symmetry all diagram have the same value and each has the same look identical to $D$ above. Lets name it $D_{\square}$. Then, in Eq. (\ref{eq:F}) $w = 24 D_\square$. The relative magnitude of coplanar and non-coplanar terms is obviously important. In Figure \ref{fig:vert1} we plotted $u(\alpha)/8$ and in Figure \ref{fig:vert2} -- $w/24$.

\section{Variational crystalline states}\label{sec:var}

\begin{figure*}
\includegraphics[width=7 in]{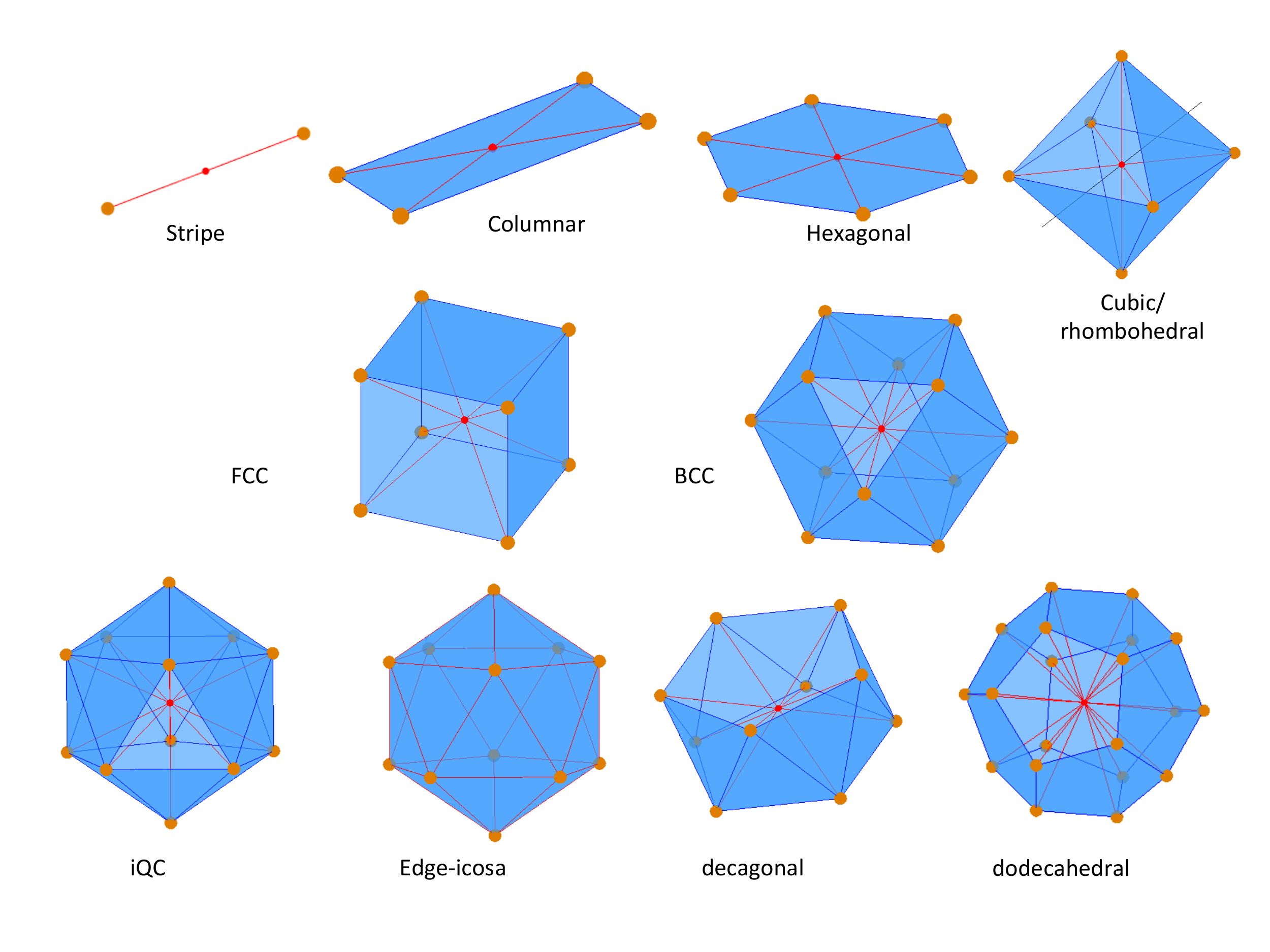}
\caption{Primary reciprocal lattice shells of variational states that we considered. Reciprocal vectors are indicated by red lines that start from red sphere and end at golden spheres. Within assumptions of Weak Crystallization theory, only such vectors, which all have equal length, contribute to crystallization energy. Clarifications: (1) vertices of perfect octahedron represent the primary reciprocal shell of a Cubic state; general rhombohedral case can be obtained by linear dilatation or contraction along the axis connecting centers of opposite triangular faces (black line). (2) Reciprocal vectors of the edge-icosahedral  state are, as the name implies, edges of icosahedron, shown in red.
}\label{fig:shell}
\end{figure*}

We have considered the following variational states ($N$ is the number of $\pm \mathbf q_i$ pairs, see Figure \ref{fig:shell}):

\begin{itemize}
  
 \item
  Smectic or stripe: \(N = 1\). 
  \item
  Columnar: \(N = 2\). 1 neighbor at optimal anlge \(\alpha_{min}\)
\item
  Rhombohedral: \(N = 3\). 2 neighbors at optimal angle \(\alpha_{min}\)
\item
BCC lattice (FCC reciprocal): \(N = 6\). 4 neighbors with
  \(\alpha = \pi/3\), 1 with \(\alpha = \pi/2\);\\
\item
  FCC lattice (BCC reciprocal): \(N = 4\). 3 neighbors with
  \(\alpha = \cos^{-1} (1/3)\)
\item
iQC: \(N = 6\). 5 neighbors with
  \(\alpha_i \approx 63.4^o\)
\item
Edge-icosahedral (momenta are the edges of icosahedron -- favored by cubic
  interaction which we neglect): \(N = 15\).  4 neighbors with
        $\alpha = 60^o$, 4 neighbors with $\alpha = 72^o$, 4 neighbors with $\alpha =
        36^o$, and 2 neighbors with $\alpha = 90^o$. In the energy there are non-coplanar terms present; we did not include this contribution since the energy of this states is relatively too high (due to many suboptimal angles $\alpha$) and the non-coplanar contribution is weighted by small factor $N^{-2}$.
\item
Decagonal (same as iQC, but with one vector pair missing): \(N = 5\). 4 neighbors at icosahedral angles $\alpha_i$.
\item
 Dodecahedral in momentum space: \(N = 10\). 3 neighbors with
  \(\alpha_1\approx 41.8^o\), 6 neighbors with
  \(\alpha_2\approx 70.5^o\). 
\item
  Hexagonal: \(N = 3\). 2 neighbors at $60^o$.
\end{itemize}

\section{Splitting peaks}\label{sec:frac}

Here we show that splitting of one Bragg peak into a pair is unfavorable. This is an immediate consequence of $u(\alpha)$ being smooth as $\alpha \to 0$, as is the case  for electron-mediated and local interactions. Indeed, assume that there is an energetically favorable (possibly multi-$\mathbf q$) configurations with a spot at $\mathbf q_0$ with amplitude $\rho_{\mathbf q_0}$. Now, suppose we split it into two at $\mathbf q_0'$ and $\mathbf q_0''$, both approximately equal to $\mathbf q_0$. To keep the interaction with the other momentum components unchanged (we assumed it to be optimal), we need $ |\rho_{\mathbf q_0'}|^2 + |\rho_{\mathbf q_0''}|^2 = |\rho_{\mathbf q_0}|^2$. That keeps the second order (r) and the interaction with distant $q$ components intact. However, instead of the original self-interaction we now have $u_0|\rho_{\mathbf q_0}|^4 \to u_0|\rho_{\mathbf q_0''}|^4 + 4 u(\alpha) |\rho_{\mathbf q_0'}|^2 |\rho_{\mathbf q_0''}|^2 \approx u_0(|\rho_{\mathbf q_0'}|^2 +|\rho_{\mathbf q_0''}|^2)^2 + 2 u(\alpha) |\rho_{\mathbf q_0'}|^2 |\rho_{\mathbf q_0''}|^2 $. Hence, the energy goes up, and splitting is not favored for $u_0>0$. Indeed, the crystallization simulations starting from random initial conditions show the extinction behavior: large Bragg peak suppresses  its smaller neighbors, leaving in the end only a small number of spots that correspond to a (q)crystal.

\section{Distorted iQC state}\label{sec:dist}

To explore the stability of iQC state with respect to distortions away from perfect icosahedron, let us expand the interaction energy in the
vicinity of the iQC:

\[E_{int} = \sum_{i\lt j} u(\alpha _{ij}) |\rho_i|^2 |\rho_j|^2.\]

For the sake of argument will neglect the fact that the amplitudes of
the order parameter can also react to distortions -- this will only
further lower the energy of distorted state. Then, defining
\(\delta_{ij} = \alpha_{ij} -\alpha_0\),

$$E_{int} =  E_0 + u'(\alpha _0)\sum_{i\lt j}\delta_{ij} +  0.5 u''(\alpha _0)\sum_{i\lt j}\delta^2_{ij}+ ...$$

Now we can define convenient coordinates for the
Bragg peaks on the sphere, and explore whether the energy can be lowered
by a distortion. Both the first and second derivate terms define
quadratic forms with non-negative eigenvalues (due to the nonlinear
dependence of \(\delta_{ij}\) on local coordinates, even the first order
term produces quadratic form upon expansion). Out of 12 total
eigenvalues, the quadratic form of \(\sum_{i\lt j}\delta_{ij}\) has only
4 non-zeros; in contrast \(\sum_{i\lt j}\delta^2_{ij}\) has only 3 zero
modes that correspond to rigid global rotations. When put together, for
\(u'(\alpha _0) \lt - (2/3) u''(\alpha _0)\) negative stiffness modes
emerge, signifying distortive instability of icosahedron. The strongest
instability occurs at the largest possible quasimomenta \(\pm 4\pi/5\)
(see Mathematica code attached to \cite{knowen}). At the critical point
\(u'(\alpha _0) = - (2/3) u''(\alpha _0)\), four zero modes
simultaneously appear, forming a flat zero-frequency band as a function
of quasimomentum on icosahedron.

Hence, the conclusion is that even if \(u(\alpha)\) reaches the minimum
at non-icosahedral anlge, the iQC remains (at least) locally stable for
\(u' \gt 0\) (``tensile strain" between Bragg peaks), and even for
``compressive strain" it remain stable until a critical value of negative
\(u'\) is reached.

\end{document}